\documentclass{IEEEtran}
\usepackage{algorithm}
\usepackage{algpseudocode}
\usepackage{times,amssymb,amsmath,amsfonts,nicefrac,float,accents,color,
graphicx,bbm,caption,subcaption,mathrsfs,amsthm,stmaryrd,soul}
\usepackage[noadjust]{cite}
\usepackage{tikz}
\usetikzlibrary{calc}
\usepackage[mathscr]{eucal}
\newcommand\remove[1]{}

\newtheorem{cnstr}{Construction}

\newtheorem{claim}{Claim}

\newcommand{\tikzmark}[1]{\tikz[overlay,remember picture] \node (#1) {};}
\newcommand{\DrawBox}[1][]{%
    \tikz[overlay,remember picture]{
    \draw[#1]
      ($(left)+(-0.4em,0.75em)$) rectangle
      ($(right)+(0.2em,-0.35em)$);}
}

\allowdisplaybreaks

\newtheorem{theorem}{Theorem}

\newtheorem{lemma}[theorem]{Lemma}

\newtheorem{corollary}[theorem]{Corollary}
\theoremstyle{remark}
\newtheorem{definition}{Definition}[section]
\theoremstyle{remark}
\newtheorem{remark}{Remark}[section]

\newcommand\ff{{\mathbb F}}
\newcommand{\ba}{\begin{array}}
\newcommand{\ea}{\end{array}}
\newcommand{\be}{\begin{equation}}
\newcommand{\ee}{\end{equation}}
\newcommand{\bea}{\begin{eqnarray}}
\newcommand{\eea}{\end{eqnarray}}

\newcommand\nc\newcommand
\nc\bfa{{\boldsymbol a}}\nc\bfA{{\boldsymbol A}}\nc\cA{{\mathcal A}}\nc\sA{{\mathscr A}}
\nc\bfb{{\boldsymbol b}}\nc\bfB{{\boldsymbol B}}\nc\cB{{\mathcal B}}\nc\sB{{\mathscr B}}
\nc\bfc{{\boldsymbol c}}\nc\bfC{{\boldsymbol C}}\nc\cC{{\mathcal C}}\nc\sC{{\mathscr C}}
\nc\bfd{{\boldsymbol d}}\nc\bfD{{\boldsymbol D}}\nc\cD{{\mathcal D}}
\nc\bfe{{\boldsymbol e}}\nc\bfE{{\boldsymbol E}}\nc\cE{{\mathcal E}}
\nc\bff{{\boldsymbol f}}\nc\bfF{{\boldsymbol F}}\nc\cF{{\mathcal F}}\nc\sF{{\mathscr F}}
\nc\bfg{{\boldsymbol g}}\nc\bfG{{\boldsymbol G}}\nc\cG{{\mathcal G}}\nc\sG{{\mathscr G}}
\nc\bfh{{\boldsymbol h}}\nc\bfH{{\boldsymbol H}}\nc\cH{{\mathcal H}}
\nc\bfi{{\boldsymbol i}}\nc\bfI{{\boldsymbol I}}\nc\cI{{\mathcal I}}\nc\sI{{\mathscr I}}
\nc\bfj{{\boldsymbol j}}\nc\bfJ{{\boldsymbol J}}\nc\cJ{{\mathcal J}}\nc\sJ{{\mathscr J}}
\nc\bfk{{\boldsymbol k}}\nc\bfK{{\boldsymbol K}}\nc\cK{{\mathcal K}}\nc\sK{{\mathscr K}}
\nc\bfl{{\boldsymbol l}}\nc\bfL{{\boldsymbol L}}\nc\cL{{\mathcal L}}
\nc\bfm{{\boldsymbol m}}\nc\bfM{{\boldsymbol M}}\nc\cM{{\mathcal M}}\nc\sM{{\mathscr M}}
\nc\bfn{{\boldsymbol n}}\nc\bfN{{\boldsymbol N}}\nc\cN{{\mathcal N}}\nc\sN{{\mathscr N}}
\nc\bfo{{\boldsymbol o}}\nc\bfO{{\boldsymbol O}}\nc\cO{{\mathcal O}}
\nc\bfp{{\boldsymbol p}}\nc\bfP{{\boldsymbol P}}\nc\cP{{\mathcal P}}\nc\eP{{\EuScriptP}}\nc\fP{{\mathfrak P}}
\nc\bfq{{\boldsymbol q}}\nc\bfQ{{\boldsymbol Q}}\nc\cQ{{\mathcal Q}}
\nc\bfr{{\boldsymbol r}}\nc\bfR{{\boldsymbol R}}\nc\cR{{\mathcal R}}\nc\sR{{\mathscr R}}
\nc\bfs{{\boldsymbol s}}\nc\bfS{{\boldsymbol S}}\nc\cS{{\mathcal S}}\nc\sS{{\mathscr S}}
\nc\bft{{\boldsymbol t}}\nc\bfT{{\boldsymbol T}}\nc\cT{{\mathcal T}}
\nc\bfu{{\boldsymbol u}}\nc\bfU{{\boldsymbol U}}\nc\cU{{\mathcal U}} \nc\sU{{\mathscr U}}
\nc\bfv{{\boldsymbol v}}\nc\bfV{{\boldsymbol V}}\nc\cV{{\mathcal V}}\nc\sV{{\mathscr V}}
\nc\bfw{{\boldsymbol w}}\nc\bfW{{\boldsymbol W}}\nc\cW{{\mathcal W}}\nc\sW{{\mathscr W}}
\nc\bfx{{\boldsymbol x}}\nc\bfX{{\boldsymbol X}}\nc\cX{{\mathcal X}}
\nc\bfy{{\boldsymbol y}}\nc\bfY{{\boldsymbol Y}}\nc\cY{{\mathcal Y}}
\nc\bfz{{\boldsymbol z}}\nc\bfZ{{\boldsymbol Z}}\nc\cZ{{\mathcal Z}}

\begin{document}

\title{Explicit constructions of optimal-access MDS codes with nearly optimal sub-packetization \thanks{The authors are with Dept. of ECE and ISR, University of Maryland, College Park, MD 20742, USA. Emails:  yeemmi@gmail.com, abarg@umd.edu. Research supported by
NSF grants CCF1422955 and CCF1618603.}}

\author{\IEEEauthorblockN{Min Ye} \hspace*{2in}
\and \IEEEauthorblockN{Alexander Barg,~\IEEEmembership{Fellow,~IEEE}}}

\maketitle

\begin{abstract} 
An $(n,k,l)$ MDS array code of length $n,$ dimension $k=n-r$ and sub-packetization $l$ is formed of $l\times n$ matrices over a finite field $F,$ with every column of the matrix stored on a separate node in the distributed storage system and viewed as a coordinate of the codeword. Repair of a failed node (recovery
of one erased column) can be performed by accessing a set of $d\le n-1$ surviving (helper) nodes. The code is said to have the optimal access property if the amount of data accessed at each of the helper nodes meets a lower bound on this quantity. For optimal-access MDS codes with $d=n-1,$ the sub-packetization $l$ satisfies the bound $l\ge r^{(k-1)/r}.$

In our previous work (\textcolor{black}{IEEE Trans. Inform. Theory, vol. 63, no.~4, 2017}), for any $n$ and $r,$ we presented an explicit construction of optimal-access MDS codes with sub-packetization $l=r^{n-1}.$ In this paper, we take up the question of reducing the sub-packetization value $l$ to make it approach the lower bound. We construct an explicit family of optimal-access codes with $l=r^{\lceil n/r\rceil},$ which differs from the optimal value by at most a factor of $r^2.$ These codes can be constructed over any finite field $F$ as long as $|F|\ge r\lceil n/r\rceil,$ and afford low-complexity encoding and decoding procedures.

We also define a version of the repair problem that bridges the context of regenerating codes and codes with locality constraints (LRC codes), \textcolor{black}{which we call} {\em group repair with optimal access}. In this variation, we assume that the set of $n=sm$ nodes is partitioned into $m$ repair groups of size $s,$ and require that the amount of accessed data for repair is the smallest possible whenever the $d=s+k-1$ helper nodes include all the other $s-1$ nodes from the same group as the failed node. For this problem, we construct a family of codes with the group optimal access property. These codes can be constructed over any field $F$ of size $|F|\ge n,$ and also afford low-complexity encoding and decoding procedures.
\end{abstract}


\section{Introduction}
The repair problem of array codes is motivated by applications of codes in distributed storage systems (e.g., the Google File System (GFS) and Hadoop Distributed File System (HDFS)) which assume that the
data is spread across a large number of drives (nodes). A popular solution of the task of protecting the data from node failures relies on maximum distance separable (MDS) array codes which provide a universal mechanism of node recovery regardless of the location of the 
failed nodes. {By distributing the codeword across different nodes, we ensure that
in the event of node failure it is possible to recover the missing data using the information stored in functional nodes. }
Among the parameters of the code that are important for storage applications are the amount of data transferred in the 
system during node repair (the repair bandwidth), which characterizes the network usage, and the volume of accessed data 
which corresponds to the number of disk I/O operations. Therefore, recent research 
on MDS codes for distributed storage has focused on codes that can minimize these two quantities; see in particular the paper
by Dimakis et al. \cite{Dimakis10} which motivated most of the recent research in coding for storage and derived lower bounds on the repair bandwidth of MDS codes.
 

\subsection{Exact-repair regenerating codes}
 Most studies of codes with optimal repair bandwidth in the literature are concerned with a particular class of 
codes known as {\em array codes} \cite{Blaum98}. 
An $(n,k,l)$ array code $\sC$ is formed of $l\times n$ matrices $(C_1,\dots,C_n)\in (F^l)^n,$ where $F$ is a finite field. Each column $C_i$ of the matrix is a codeword coordinate, and the parameter $l$ that determines the dimension of the column vector $C_i$ is called {\em sub-packetization}.
The code $\sC$ is said to have the MDS property if every $k$ out of $n$ columns of the matrix suffice to recover the remaining $r$ columns. In this paper, we consider only MDS array codes. 
As usual in distributed storage applications, we assume that an MDS array code of length $n$ formed of $k$ information coordinates and 
$r=n-k$ parity coordinates is spread across $n$ different nodes of the storage cluster. Each node of the cluster stores a coordinate 
of the code. 
At the same time, the basic repair task studied for MDS array codes consists in recovering one erased column (a failed node) by accessing information stored in the other nodes of the same codeword.

Suppose that a node becomes unavailable, and the system attempts to repair its content by connecting to $d$ surviving (helper) 
nodes, $k\le d\le n-1$. From the perspective of system architecture, efficient repair requires that the amount of information accessed and downloaded from the helper nodes be as small as possible, and this introduces the notion of {access} and {repair bandwidth}. 
 To formalize this concept, let us give the following definition.
%
\begin{definition}\label{def:bandwidth} Consider an $(n,k,l)$ MDS array code $\sC$ over $F$.
\textcolor{black}{We write a codeword of $\cC$ as $(C_1,\dots,C_n)$,}
where $C_i=(c_{i,0},c_{i,1},\dots,c_{i,l-1})^T\in F^l, i=1,\dots, n$. 
We say that a node $i\in [n]$ can be repaired from a subset of helper nodes $\sR_i\subset[n]\backslash \{i\}, |\sR_i|\ge k$
by accessing $\omega_i(\sR_i)$ symbols of $F$ and downloading $\beta_i(\sR_i)$ symbols of $F$ 
if there are
\textcolor{black}{ numbers $\beta_{i,j}, j\in\sR_i$ and $\omega_{i,j}, j\in\sR_i$, and $|\sR_i|+1$
 functions $f_{i,j}: F^l\to F^{\beta_{i,j}}, j\in\sR_i$
and $g_i: F^{\sum_j\beta_{i,j}}\to F^l$ such that}
   \begin{align*}
   & C_i=g_i(f_{i,j}(C_{j}),j\in\sR_i ), \\
& \sum_{j\in\sR_i}\beta_{i,j}=\beta_i(\sR_i), \text{ and }
\sum_{j\in\sR_i}\omega_{i,j}=\omega_i(\sR_i),
   \end{align*}
where the function $f_{i,j}$ depends on $\omega_{i,j}$ coordinates of the node $C_j=(c_{j,0},c_{j,1},\dots,c_{j,l-1})^T.$

\textcolor{black}{Informally, this definition says that we download some functions $f_{i,j}$ of the information stored in the helper nodes and perform repair using the function $g_i$ that takes the values of $f_{i,j}$ as its arguments. The quantities $\omega_i(\sR_i)$ and $\beta_i(\sR_i)$ control the
number of the accessed and downloaded symbols of the field $F$, respectively.}

The smallest volume of the downloaded data $\beta^\ast_i(\sR_i)=\min \beta_i(\sR_i)$ over the choice of the functions $\{f_{i,j}\}_{j\in\sR_i}$ and $g_i$
is called the $(i,\sR_i)$-\emph{repair bandwidth} of the code $\sC.$
The quantity $\beta(\sC):=\max_{i\in[n]}\beta^\ast_i([n]\backslash\{i\})$ is called the \emph{repair bandwidth} of the code $\sC.$

The smallest amount of the accessed data $\omega^\ast_i(\sR_i)=\min \omega_i(\sR_i)$ over the choice of the functions $\{f_{i,j}\}_{j\in\sR_i}$ and $g_i$
is called the $(i,\sR_i)$-\emph{access} of the code $\sC.$
The quantity $\omega(\sC):=\max_{i\in[n]}\omega^\ast_i([n]\backslash\{i\})$ is called the \emph{access} of the code $\sC.$
 Clearly, $\omega^\ast_i(\sR_i)\ge \beta^\ast_i(\sR_i)$ for any $i\in[n]$ and any $\sR_i\subset[n]\backslash \{i\}, |\sR_i|\ge k.$ Consequently, $\omega(\sC)\ge\beta(\sC)$.
\end{definition}
\vspace*{.1in}
The lower bounds on the repair bandwidth and access were established in the recent literature on regenerating codes. As shown in \cite{Dimakis10}, for an $(n,k,l)$ MDS array code $\sC,$ the recovery of a single failed node from $d$ helper nodes requires to download
at least a $1/(d+1-k)$ fraction of the data stored in each of the helper nodes. {Using our notation},
   \begin{equation}\label{eq:lowerb}
  \omega^\ast_i(\sR_i)\ge \beta^\ast_i(\sR_i)\ge \frac{|\sR_i|l}{|\sR_i|+1-k}
  \end{equation}
for any $i\in[n]$ and any $\sR_i\subset[n]\backslash \{i\}, |\sR_i|\ge k.$
Note that the right-hand side of \eqref{eq:lowerb} is a decreasing function of $|\sR_i|$.
Since \eqref{eq:lowerb} is an achievable lower bound (as discussed below in the introduction), the repair bandwidth is minimized when  $\sR_i = [n]\backslash \{i\}$. In this case, \eqref{eq:lowerb} becomes
\begin{equation}\label{eq:orc}
\omega^\ast_i( [n]\backslash \{i\} )\ge \beta^\ast_i( [n]\backslash \{i\} )\ge 
\frac{(n-1) l}{n-k}.
\end{equation}


We say that an $(n,k,l)$ MDS array code $\sC$ has the {\em optimal repair} property
(and call it an optimal-repair $(n,k,l)$ code) if $\beta(\sC)=\frac{(n-1)l}{n-k}.$ A coding-theoretic perspective of the bound \eqref{eq:lowerb} and of other related results was recently developed in \cite{CM15}.

According to Definition \ref{def:bandwidth}, even for codes with the optimal repair property, we might still need to access a larger amount of data than the lower bounds in  \eqref{eq:lowerb} during the recovery of a failed node. We say that an $(n,k,l)$ MDS array code $\sC$ has the {\em optimal access} property
(and call it an optimal-access $(n,k,l)$ code) if 
$\omega(\sC)=\frac{(n-1)l}{n-k}.$
We call $\sC$ a \emph{systematic optimal-access} MDS array code if $\omega^\ast_i([n]\backslash\{i\})=\frac{(n-1)l}{n-k}$ for any systematic node $C_i.$
 Clearly, optimal-access MDS array codes are a subclass of optimal-repair MDS array codes.

The construction problem of optimal-repair MDS array codes has been extensively studied over the last several years.
More specifically, for $k\le (n+1)/2$ (the low rate regime), MDS array codes with $d$-optimal repair property were constructed in 
\cite{Rashmi11,Shah12,Wu09}.
For $k>(n+1)/2$ (the high-rate regime)
papers \cite{Cadambe11,Cadambe11a,Papai13,Tamo13,Wang14} showed that for large enough base field $F$ there exist MDS array codes that can 
optimally repair any single systematic node failure using all the surviving nodes, and \cite{Wang11} showed the same for all rather than 
only systematic nodes.
However, explicit constructions of such codes with rate larger than $1/2$ and $r>3$ were found only recently \cite{Ye16}, where
we proposed two families of MDS array codes that can repair any number of failed nodes from any number of helper nodes by downloading the minimum possible amount of data from the helper nodes.

The subclass of optimal-access codes is of particular interest among optimal-repair codes. At the same time, 
the sub-packetization parameter of optimal-access codes is constrained from below. Namely, 
according to a result of \cite{Tamo14}, if an $(n,k,l)$ MDS array code has the optimal access property, then
$l\ge r^{(k-1)/r}.$ Existence and constructions of optimal-access codes were studied in several recent works.
We mention the results of \cite{Wang11} which established existence of optimal-access MDS array codes with $l=r^k$ and
\cite{Sasid15,Rawat16} which proved existence of such codes with $l=r^{\lceil n/r\rceil}$, although both results require a large-size
finite field $F.$ As for explicit constructions, until recently they were known only for $r=2,3$. 
\textcolor{black}{Namely, \cite{Wang11}
constructed codes with $l=r^k$ over the field $\ff_3$ if $r=2$ and $\ff_4$ if $r=3$ (independent of $k$).
The paper \cite{Raviv15} constructed systematic optimal-access 
MDS array codes\footnote{\textcolor{black}{\cite{Raviv15} also considered relaxing systematic optimal-access to systematic optimal-repair, and gave an explicit construction of codes for $r=3$ and $k$ a multiple of $4$ with sub-packetization $l=r^{k/(r+1)}$  over the field of size linear in $k.$}}
 with $l=r^{k/r},$ where $k$ is a 
multiple of $r$. The size of the underlying field for \cite{Raviv15} is at least $k/2+1$ for $r=2$ and $2k+1$ for $r=3$.}
 In \cite {Ye16}, we proposed a family of optimal-access MDS array codes with sub-packetization
$l=r^{n-1},$ and this is the only explicit family of optimal-access MDS array codes for $r>3$ known in the literature.

In this work we present an explicit construction of optimal-access MDS array codes for any $r$ and $n$ with sub-packetization $l=r^{\lceil n/r\rceil},$ which differs from the lower bound by a factor of at most $r^2.$ These codes can be constructed over any finite field $F$ as long as $|F|\ge r\lceil n/r\rceil,$ and the encoding and decoding procedures of these codes have low complexity.

\begin{remark} The repair problem has been studied for a relatively short time, so the related terminology is still somewhat unsettled. For instance, \cite{Sasid15,Rawat16} 
refer to codes with $l=r^{\lceil n/r\rceil}$ as codes with polynomial sub-packetization. This implicitly assumes the
asymptotic regime of $k=Rn,$ i.e., of codes with a fixed rate $R$ bounded away from 1. At the same time, arguably the case
of $r=o(n),$ for instance constant $r,$ is more important for the repair problem because the encoded data in storage are likely to include only a small
number of parity checks. In this regime the above value of $l$ is an exponential function of the block length. To cover all the possible cases, we prefer not to use the terms polynomial or exponential to describe the growth
rate of the parameter $l.$
\end{remark}

\begin{remark}
{Recently, the repair problem was extended from array codes to classes of scalar codes, such as Reed-Solomon (RS) codes. This line of research was initiated by Guruswami and Wootters \cite{Guruswami17} who gave a characterization of linear repair schemes of general scalar linear MDS codes. They also
presented a specific repair scheme for a family of RS codes and proved that (in some cases) the
repair bandwidth of RS codes under this scheme is the smallest possible among all linear repair schemes
and all scalar linear MDS codes with the same parameters.
At the same time, the repair bandwidth of RS codes attained in \cite{Guruswami17} is rather far from the bound \eqref{eq:orc}. 

In a subsequent work \cite{Ye16a}, the authors used the approach of \cite{Guruswami17} to construct an explicit family of RS codes whose repair bandwidth asymptotically achieves the bound \eqref{eq:orc}.
Very recently, I. Tamo and the present authors \cite{Tamo17} presented an explicit construction of RS codes that achieve the bound \eqref{eq:orc} for any given code parameters $n$ and $k$. While the construction
in \cite{Tamo17} has superexponential sub-packetization (specifically, the sub-packetization scales as $e^{n\log n}$), the same paper shows this growth order is also necessary for scalar linear codes to achieve the bound \eqref{eq:orc} using linear repair schemes.
In contrast to this, for vector (array) codes to attain the bound \eqref{eq:orc}, it is sufficient
to use exponentially growing sub-packetization.}

\end{remark}

\begin{remark} {\sc Added on June 30, 2017:} Shortly after the release of this paper, in an independent work, B. Sasidharan et al. \cite{Sasidharan16} presented a construction of codes that is very similar to our Construction~\ref{construction:cns}. 
\end{remark}

\subsection{Repair groups and node regeneration: Group optimal access property}
In this paper, we also introduce a coding problem for distributed storage 
that bridges codes with locality, in particular, LRC codes \textcolor{black}{(e.g., \cite{Papai14,Tamo13a,Silb13})} and regenerating codes. To motivate it, 
consider an architecture of distributed storage under which $n=sm$
storage nodes are partitioned into $m$ local groups \textcolor{black}{(we assume throughout that $s\le r$)}. Nodes in the same group are logically better connected (for instance, they are 
geographically close to each other and thus have stable links between them), while the connectivity between nodes from 
different groups fluctuates smoothly over time (for instance, relying on a slowly fading channel).
When a node fails, the system seeks to perform repair by accessing the minimum possible amount of data on a set of helper nodes. Efficient repair also suggests that the helper nodes are chosen from a subset of nodes most easily reachable from the failed node. Since the failed node can always connect to all the nodes in the same group as itself,
an MDS array code with the $(s,d=s+k-1)$-group optimal access property can minimize the disk I/O and network traffic during the repair of any single failed node for such systems.

This motivates the following definition.
\begin{definition} Let $\sC$ be an $(n=sm,k,l)$ MDS array code whose nodes are partitioned into $m$ groups of size $s$ each.
Referring to Def.~\ref{def:bandwidth}, we say that $\sC$ has the \emph{$(s,d=s+k-1)$-group optimal access property} if $ \omega^\ast_i(\sR_i)=dl/s$ for any $i$ and any set  of helper nodes $\sR_i$ of size $d$ that contains all the $s-1$ nodes from the same group as $i.$
\end{definition}
By definition, repair of the failed node in the group repair mode can be performed by accessing the volume of data that attains the lower bound \eqref{eq:lowerb}, justifying the optimality qualifier.

In this paper, we construct an  explicit family of $(n=sm,k=n-r,l=s^m)$ MDS array codes
with the $(s,d=s+k-1)$-group optimal access property for any $s,m$ and $r$ such that $r\ge s.$
These codes can be constructed over any finite field $F$ as long as $|F|\ge n,$ and are equipped with low-complexity encoding and decoding. Our construction is flexible in the sense that it allows any number $m$ of local groups and any number \textcolor{black}{$s \le r$} of storage nodes in a local group. 

The code construction is presented in the next section. Section \ref{sect:MDS} contains a proof of the MDS property of the constructed codes, and Section \ref{sect:opac} gives a proof of the group optimal access property.

We note that coding designs that address data regeneration based on different conditions at different helper nodes, based on
access conditions or transient unavailability (degraded reads or hard errors) have been considered in a number of earlier works.
For instance, in \cite{Ye16} we constructed codes that support repair of one or more failed nodes by accessing any set of $d$ helper nodes in the same encoding block. A different line of research
 that establishes conditions under which helper node selection improves the
storage/bandwidth tradeoff was recently developed in \cite{Ahmad16}. Yet another link between regenerating codes and LRC codes is the ``local regeneration" problem \cite{kamath14,hollmann14},
where the local repair of the code is also required to have small bandwidth.

\section{Code construction}\label{sect:cns}

Let $\sC\in F^{ln}$ be an $(n,k,l)$ array code with nodes $C_i\in F^l, i=1,\dots,n,$ where each $C_i$ is a column vector with coordinates $C_i=(c_{i,0}, c_{i,1},\dots, c_{i,l-1})^T.$ 
Throughout this paper we consider codes defined by the following $r$ parity-check equations:
\begin{equation}\label{eq:parityform}
\sC=\{(C_1,C_2,\dots,C_n):\sum_{i=1}^n A_{t,i}C_i=0,\, t=0,1,\dots,r-1\},
\end{equation}
where $A_{t,i},0\le t\le r-1, 1\le i\le n$ are $l\times l$ matrices over $F$.  
Let $A(a,b)$ be the entry in the $a$-th row and $b$-th column of matrix $A,$ $0\le a,b\le l-1.$ Throughout we assume that $0^0=1.$

\begin{cnstr}\label{construction:cns} Let $s,r$ and $m$ be positive integers such that $s\le r\le sm,$ let $n=sm, l=s^m.$
Let $F$ be a finite field of size $|F|\ge n,$ let $\{\lambda_i\}_{i\in[n]}$ be $n$ distinct elements in $F,$ and let
$\gamma\in F\backslash\{0,1\}.$  Given an integer $a, 0\le a\le l-1,$ we write its $s$-ary expansion as 
$
a=(a_m,a_{m-1},\dots,a_1).
$

For $v\in[m]$ and $0\le u\le s-1,$ define $a(v,u):=(a_m,a_{m-1},\dots,a_{v+1},u,a_{v-1},a_{v-2}\dots,a_1).$
For $v\in[m],u=0,1,\dots,s-1,$ and $t=0,1,\dots, r-1,$ define an $l\times l$ matrix $A_{t,(v-1)s+u+1}$ as follows: for $0\le a,b\le l-1,$ let
\begin{equation}\label{eq:osp}
\begin{aligned}
& A_{t,(v-1)s+u+1}(a,b)=\\
& \left\{ \begin{array}{ll}
\lambda_{(v-1)s+u+1}^t & \mbox{if }  a_v< u,b=a, \\ 
\gamma\lambda_{(v-1)s+u+1}^t & \mbox{if }  a_v>u,b=a, \\ 
\lambda_{(v-1)s+w+1}^t & \mbox{if }  a_v=u,b=a(v,w) \\
 &  \text{ for some }w\in\{0,1,\dots,s-1\}, \\ 
0 & \mbox{otherwise.} \end{array}\right. 
\end{aligned}
\end{equation}
We construct an $(n,k=n-r,l)$ array code defined by \eqref{eq:parityform}, where the matrices $A_{t,i}$ are defined in 
\eqref{eq:osp}.
\end{cnstr}

We will show that the code $\sC$ defined by Construction \ref{construction:cns} has the MDS property. In the case of $s=r$ it also 
has the optimal access property, while if $s< r$ it has the group optimal access property.

In Section~\ref{sect:MDS} we give an example of the above matrices for $s=r=3$ and $m=2$ and show that the obtained codes 
have the MDS property.

In Construction \ref{construction:cns} we assumed that the code length is a product of two numbers, $s$ and $m$. While this assumption 
leads to a simple uniform formulation of the code construction, it can be easily lifted at the expense of a more detailed notation.
Namely, the following construction extends the case of $s=r$ in Construction \ref{construction:cns} to cover all possible code length $n$ and has essentially the same properties as the codes defined above.

\begin{cnstr}\label{construction:ext} Let $n=rm+r'$ and $l=r^{m+1}$, where $r>0,m\ge0$ are integers and $1\le r'\le r-1.$
 Let $F$ be a finite field of size $|F|\ge r(m+1),$ let $\{\lambda_i, i=1,\dots,r(m+1)\}$ be  distinct elements of $F,$ and 
let $\gamma\in F\backslash\{0,1\}.$ 
Given an integer $a$ between $0$ and $l-1,$ we write its $r$-ary expansion as 
$
a=(a_{m+1},a_m,\dots,a_1).
$
For $v\in\{1,\dots,m+1\}$ and $0\le u\le r-1,$ define $a(v,u):=(a_{m+1},a_m,\dots,a_{v+1},u,a_{v-1},a_{v-2}\dots,a_1).$

For $v\in[m],u=0,1,\dots,r-1,$ and $t=0,1,\dots, r-1,$ define an $l\times l$ matrix $A_{t,(v-1)r+u+1}$ as follows: for $0\le a,b\le l-1,$ let
    \begin{equation}\label{eq:M1}
\begin{aligned}
& A_{t,(v-1)r+u+1}(a,b)= \\
& \left\{ \begin{array}{rl}
\lambda_{(v-1)r+u+1}^t & \mbox{if }  a_v< u,b=a, \\ 
\gamma\lambda_{(v-1)r+u+1}^t & \mbox{if }  a_v>u,b=a, \\ 
\lambda_{(v-1)r+w+1}^t & \mbox{if }  a_v=u,b=a(v,w) \\
 &  \text{ for some }w\in\{0,1,\dots,r-1\}, \\ 
0 & \mbox{otherwise.} \end{array}\right. 
\end{aligned}
    \end{equation}
For $u=0,1,\dots,r'-1,$ and $t=0,1,\dots, r-1,$ define an $l\times l$ matrix $A_{t,mr+u+1}$ as follows:
for $0\le a,b\le l-1,$ let
   \begin{equation}\label{eq:M2}
\begin{aligned}
& A_{t,mr+u+1}(a,b)= \\
& \left\{ \begin{array}{rl}
\lambda_{mr+u+1}^t & \mbox{if }  a_{m+1}< u,b=a, \\ 
\gamma\lambda_{mr+u+1}^t & \mbox{if }  a_{m+1}>u,b=a, \\ 
\lambda_{mr+w+1}^t & \mbox{if }  a_{m+1}=u,b=a(m+1,w) \\
 & \text{ for some }w\in\{0,1,\dots,r-1\}, \\ 
0 & \mbox{otherwise.} \end{array}\right. 
\end{aligned}
   \end{equation}
We construct an $(n=rm+r',k=n-r,l=r^{m+1})$ array code defined by \eqref{eq:parityform}, where the matrices $A_{t,i},0\le t\le r-1, 1\le i\le n$ are defined in \eqref{eq:M1}-\eqref{eq:M2}.
\end{cnstr}

\section{The MDS property}\label{sect:MDS}

In this section we show that the code family given by Construction \ref{construction:cns} has the MDS property. We start with
an example that shows the working of the definition \eqref{eq:parityform}-\eqref{eq:osp}
as well as provides intuition for the proof of the MDS property given below in this section. While the notation in the proof
makes it difficult to glean an intuitive picture, this example serves to visualize the ideas behind the construction and the proof.

\subsection{Example}\label{subsect:example}
Take $s=r=3$ and $m=2$ in Construction \ref{construction:cns}, so $n=6$ and $l=9.$ Let us write out the $9\times 9$ matrices $A_{t,i},i=1,\dots, 6.$ The 
code presented below can be realized over any field of size $|F|\ge n=6,$ so the smallest field is $\ff_7.$
\vspace*{.1in}
{\footnotesize
$$
A_{t,1}={\left[ \begin{array}{*{9}{c}} 
\lambda_1^t & \lambda_2^t & \lambda_3^t & 0 & 0 & 0 & 0 & 0 & 0\\
0 & \gamma\lambda_1^t & 0 & 0 & 0 & 0 & 0 & 0 & 0\\
0 &  0 & \gamma\lambda_1^t & 0 & 0 & 0 & 0 & 0 & 0\\
 0 & 0 & 0 & \lambda_1^t & \lambda_2^t & \lambda_3^t & 0 & 0 & 0\\
 0 & 0 & 0 & 0 & \gamma\lambda_1^t & 0 & 0 & 0 & 0\\
 0 & 0 & 0 & 0 &  0 & \gamma\lambda_1^t & 0 & 0 & 0\\
 0 & 0 & 0 &  0 & 0 & 0 & \lambda_1^t & \lambda_2^t & \lambda_3^t \\
 0 & 0 & 0 &  0 & 0 & 0 & 0 & \gamma\lambda_1^t & 0 \\
 0 & 0 & 0 &  0 & 0 & 0 &0 &  0 & \gamma\lambda_1^t \end{array} \right]}
$$\par} 

{\footnotesize
$$
A_{t,2}={\left[ \begin{array}{*{9}{c}} 
\lambda_2^t & 0 & 0 & 0 & 0 & 0 & 0 & 0 & 0\\
\lambda_1^t & \lambda_2^t & \lambda_3^t & 0 & 0 & 0 & 0 & 0 & 0\\
0 &  0 & \gamma\lambda_2^t & 0 & 0 & 0 & 0 & 0 & 0\\
 0 & 0 & 0 & \lambda_2^t & 0 & 0 & 0 & 0 & 0\\
 0 & 0 & 0 & \lambda_1^t & \lambda_2^t & \lambda_3^t & 0 & 0 & 0\\
 0 & 0 & 0 & 0 &  0 & \gamma\lambda_2^t & 0 & 0 & 0\\
 0 & 0 & 0 &  0 & 0 & 0 & \lambda_2^t & 0 & 0 \\
 0 & 0 & 0 &  0 & 0 & 0 & \lambda_1^t & \lambda_2^t & \lambda_3^t \\
 0 & 0 & 0 &  0 & 0 & 0 &0 &  0 & \gamma\lambda_2^t \end{array} \right]}
$$\par}

{\footnotesize
$$
A_{t,3}=\left[ \begin{array}{*{9}{c}} 
\lambda_3^t & 0 & 0 & 0 & 0 & 0 & 0 & 0 & 0\\
0 & \lambda_3^t & 0 & 0 & 0 & 0 & 0 & 0 & 0\\
\lambda_1^t & \lambda_2^t & \lambda_3^t & 0 & 0 & 0 & 0 & 0 & 0\\
 0 & 0 & 0 & \lambda_3^t & 0 & 0 & 0 & 0 & 0\\
 0 & 0 & 0 & 0 & \lambda_3^t & 0 & 0 & 0 & 0\\
 0 & 0 & 0 & \lambda_1^t & \lambda_2^t & \lambda_3^t & 0 & 0 & 0\\
 0 & 0 & 0 &  0 & 0 & 0 & \lambda_3^t & 0 & 0 \\
 0 & 0 & 0 &  0 & 0 & 0 &0 & \lambda_3^t & 0\\
 0 & 0 & 0 &  0 & 0 & 0 & \lambda_1^t & \lambda_2^t & \lambda_3^t  \end{array} \right]
$$}

{\footnotesize
$$
A_{t,4}=\left[ \begin{array}{*{9}{c}} 
\lambda_4^t & 0 & 0 & \lambda_5^t & 0 & 0 & \lambda_6^t & 0 & 0\\
 0 & \lambda_4^t & 0 & 0 & \lambda_5^t & 0 & 0 & \lambda_6^t & 0\\
 0 & 0 & \lambda_4^t &  0 & 0 & \lambda_5^t & 0 & 0 & \lambda_6^t \\
0 & 0 & 0 &  \gamma\lambda_4^t & 0 & 0 & 0 & 0 & 0\\
 0 & 0 & 0 & 0 & \gamma\lambda_4^t & 0 & 0 & 0 & 0\\
 0 & 0 & 0 &  0 & 0 & \gamma\lambda_4^t & 0 & 0 & 0 \\
0 &  0 & 0 & 0 & 0 & 0 & \gamma\lambda_4^t & 0 & 0\\
 0 & 0 & 0 & 0 &  0 & 0 & 0 & \gamma\lambda_4^t & 0\\
 0 & 0 & 0 &  0 & 0 & 0 &0 &  0 & \gamma\lambda_4^t \end{array} \right]
$$}

{\footnotesize
$$
A_{t,5}=\left[ \begin{array}{*{9}{c}} 
\lambda_5^t & 0 & 0 & 0 & 0 & 0 & 0 & 0 & 0\\
 0 & \lambda_5^t & 0 & 0 & 0 & 0 & 0 & 0 & 0\\
 0 & 0 & \lambda_5^t &  0 & 0 & 0 & 0 & 0 & 0 \\
\lambda_4^t & 0 & 0 & \lambda_5^t & 0 & 0 & \lambda_6^t & 0 & 0\\
 0 & \lambda_4^t & 0 & 0 & \lambda_5^t & 0 & 0 & \lambda_6^t & 0\\
 0 & 0 & \lambda_4^t &  0 & 0 & \lambda_5^t & 0 & 0 & \lambda_6^t \\
0 &  0 & 0 & 0 & 0 & 0 & \gamma\lambda_5^t & 0 & 0\\
 0 & 0 & 0 & 0 &  0 & 0 & 0 & \gamma\lambda_5^t & 0\\
 0 & 0 & 0 &  0 & 0 & 0 &0 &  0 & \gamma\lambda_5^t \end{array} \right]
$$}

{\footnotesize
$$
A_{t,6}=\left[ \begin{array}{*{9}{c}} 
\lambda_6^t & 0 & 0 & 0 & 0 & 0 & 0 & 0 & 0\\
 0 & \lambda_6^t & 0 & 0 & 0 & 0 & 0 & 0 & 0\\
 0 & 0 & \lambda_6^t &  0 & 0 & 0 & 0 & 0 & 0 \\
0 & 0 & 0 & \lambda_6^t & 0 & 0 & 0 & 0 & 0\\
 0 & 0 & 0 & 0 & \lambda_6^t & 0 & 0 & 0 & 0\\
 0 & 0 & 0 &  0 & 0 & \lambda_6^t &0 & 0 & 0\\
\lambda_4^t & 0 & 0 & \lambda_5^t & 0 & 0 & \lambda_6^t & 0 & 0\\
 0 & \lambda_4^t & 0 & 0 & \lambda_5^t & 0 & 0 & \lambda_6^t & 0\\
 0 & 0 & \lambda_4^t &  0 & 0 & \lambda_5^t & 0 & 0 & \lambda_6^t  \end{array} \right]
$$}

The MDS property states that any $3\times 3$ block submatrix of the $3\times 6$ block matrix formed of the matrices $A_{t,i}$ 
is invertible (here the blocks are $l\times l$ matrices). Below we show this for the matrix
   $$
B=
\left[\begin{array}{ccc}
A_{0,1} & A_{0,2} & A_{0,5}\\
A_{1,1} & A_{1,2} & A_{1,5}\\
A_{2,1} & A_{2,2} & A_{2,5}
\end{array}\right].
   $$
Let $X$ is a column vector in $F^{27}$ with coordinates $X=(x_0, x_1, \dots, x_{26})^T.$ Our claim will follow if we prove that
$BX=0$ implies that $X=0.$ 

We proceed as follows. For convenience of presentation, let us permute the rows of $B$ to obtain a matrix $D=PB$, where the permutation matrix $(P_{ij})_{0\le i,j\le 26}$ is given by
  \begin{equation}\label{eq:mod9}
    P_{ij}=1 \text{ iff } i=(j-j\,\text{mod\,}9)/9+3 (j\,\text{mod\,}9).
  \end{equation}
It is clear that $P$ has exactly one 1 in each row, so it is indeed a permutation on $\{0,1,\dots,26\}$ (note that multiplication by a full-rank matrix does not change the rank). 
We will prove that the matrix $D$ has a trivial null space, i.e., $DX=0$ implies that $X=0.$
Writing out the condition $DX=0$ explicitly, we obtain a system of equations given in \eqref{eq:huge}.

\begin{figure*}
\begin{equation}\label{eq:huge}
\begin{aligned}
\footnotesize{
\left[ \begin{array}{*{27}{@{\hspace*{.03in}}c}} 
1 & 1 & 1 & 0 & 0 & 0 & 0 & 0 & 0 & 1 & 0 & 0 & 0 & 0 & 0 & 0 & 0 & 0 & 1 & 0 & 0 & 0 & 0 & 0 & 0 & 0 & 0\\
\lambda_1 & \lambda_2 & \lambda_3 & 0 & 0 & 0 & 0 & 0 & 0 & \lambda_2 & 0 & 0 & 0 & 0 & 0 & 0 & 0 & 0 & \lambda_5 & 0 & 0 & 0 & 0 & 0 & 0 & 0 & 0\\
\lambda_1^2 & \lambda_2^2 & \lambda_3^2 & 0 & 0 & 0 & 0 & 0 & 0 & \lambda_2^2 & 0 & 0 & 0 & 0 & 0 & 0 & 0 & 0 & \lambda_5^2 & 0 & 0 & 0 & 0 & 0 & 0 & 0 & 0\\
0 & \gamma & 0 & 0 & 0 & 0 & 0 & 0 & 0 & 1 & 1 & 1 & 0 & 0 & 0 & 0 & 0 & 0 & 0 & 1 & 0 & 0 & 0 & 0 & 0 & 0 & 0\\
0 & \gamma\lambda_1 & 0 & 0 & 0 & 0 & 0 & 0 & 0 & \lambda_1 & \lambda_2 & \lambda_3 & 0 & 0 & 0 & 0 & 0 & 0 & 0 &  \lambda_5 & 0 & 0 & 0 & 0 & 0 & 0 & 0\\
0 & \gamma\lambda_1^2 & 0 & 0 & 0 & 0 & 0 & 0 & 0 & \lambda_1^2 & \lambda_2^2 & \lambda_3^2 & 0 & 0 & 0 & 0 & 0 & 0 & 0 & \lambda_5^2 & 0 & 0 & 0 & 0 & 0 & 0 & 0\\
\tikzmark{left}0 &  0 & \gamma & 0 & 0 & 0 & 0 & 0 & 0 & 0 &  0 & \gamma & 0 & 0 & 0 & 0 & 0 & 0 &  0 & 0 & 1 &  0 & 0 & 0 & 0 & 0 & 0 \\
0 &  0 & \gamma\lambda_1 & 0 & 0 & 0 & 0 & 0 & 0 & 0 &  0 & \gamma\lambda_2 & 0 & 0 & 0 & 0 & 0 & 0 &  0 & 0 & \lambda_5 &  0 & 0 & 0 & 0 & 0 & 0 \\
0 &  0 & \gamma\lambda_1^2 & 0 & 0 & 0 & 0 & 0 & 0 & 0 &  0 & \gamma\lambda_2^2 & 0 & 0 & 0 & 0 & 0 & 0 &  0 & 0 & \lambda_5^2 &  0 & 0 & 0 & 0 & 0 & 0 \tikzmark{right} \\
\DrawBox[black]
 0 & 0 & 0 & 1 & 1 & 1 & 0 & 0 & 0 &  0 & 0 & 0 & 1 & 0 & 0 & 0 & 0 & 0 &1 & 0 & 0 & 1 & 0 & 0 & 1 & 0 & 0\\
 0 & 0 & 0 & \lambda_1 & \lambda_2 & \lambda_3 & 0 & 0 & 0 &  0 & 0 & 0 & \lambda_2 & 0 & 0 & 0 & 0 & 0 & \lambda_4 & 0 & 0 &  \lambda_5 & 0 & 0 & \lambda_6 & 0 & 0\\
 0 & 0 & 0 & \lambda_1^2 & \lambda_2^2 & \lambda_3^2 & 0 & 0 & 0 &  0 & 0 & 0 & \lambda_2^2 & 0 & 0 & 0 & 0 & 0 & \lambda_4^2 & 0 & 0 & \lambda_5^2 & 0 & 0 & \lambda_6^2 & 0 & 0\\
 0 & 0 & 0 & 0 & \gamma & 0 & 0 & 0 & 0 &  0 & 0 & 0 & 1 & 1 & 1 & 0 & 0 & 0 &  0 & 1 & 0 & 0 & 1 & 0 & 0 & 1 & 0\\
 0 & 0 & 0 & 0 & \gamma\lambda_1 & 0 & 0 & 0 & 0 &  0 & 0 & 0 & \lambda_1 & \lambda_2 & \lambda_3 & 0 & 0 & 0 &  0 &  \lambda_4 & 0 & 0 & \lambda_5 & 0 & 0 & \lambda_6 & 0\\
 0 & 0 & 0 & 0 & \gamma\lambda_1^2 & 0 & 0 & 0 & 0 &  0 & 0 & 0 & \lambda_1^2 & \lambda_2^2 & \lambda_3^2 & 0 & 0 & 0 &  0 & \lambda_4^2 & 0 & 0 & \lambda_5^2 & 0 & 0 & \lambda_6^2 & 0\\
 0 & 0 & 0 & 0 &  0 & \gamma & 0 & 0 & 0 &  0 & 0 & 0 & 0 &  0 & \gamma & 0 & 0 & 0 &  0 & 0 & 1 &  0 & 0 & 1 & 0 & 0 & 1 \\
 0 & 0 & 0 & 0 &  0 & \gamma\lambda_1 & 0 & 0 & 0 &  0 & 0 & 0 & 0 &  0 & \gamma\lambda_2 & 0 & 0 & 0 &  0 & 0 & \lambda_4 &  0 & 0 &  \lambda_5 & 0 & 0 & \lambda_6 \\
 0 & 0 & 0 & 0 &  0 & \gamma\lambda_1^2 & 0 & 0 & 0 &  0 & 0 & 0 & 0 &  0 & \gamma\lambda_2^2 & 0 & 0 & 0 &  0 & 0 & \lambda_4^2 &  0 & 0 & \lambda_5^2 & 0 & 0 & \lambda_6^2 \\
 0 & 0 & 0 &  0 & 0 & 0 & 1 & 1 & 1 &  0 & 0 & 0 &  0 & 0 & 0 & 1 & 0 & 0 & 0 &  0 & 0 & 0 & 0 & 0 & \gamma & 0 & 0\\
 0 & 0 & 0 &  0 & 0 & 0 & \lambda_1 & \lambda_2 & \lambda_3 &  0 & 0 & 0 &  0 & 0 & 0 & \lambda_2 & 0 & 0 & 0 &  0 & 0 & 0 & 0 & 0 & \gamma\lambda_5 & 0 & 0\\
 0 & 0 & 0 &  0 & 0 & 0 & \lambda_1^2 & \lambda_2^2 & \lambda_3^2 &  0 & 0 & 0 &  0 & 0 & 0 & \lambda_2^2 & 0 & 0 & 0 &  0 & 0 & 0 & 0 & 0 & \gamma\lambda_5^2 & 0 & 0\\
 0 & 0 & 0 &  0 & 0 & 0 & 0 & \gamma & 0 &  0 & 0 & 0 &  0 & 0 & 0 & 1 & 1 & 1 & 0 & 0 & 0 & 0 &  0 & 0 & 0 & \gamma & 0\\
 0 & 0 & 0 &  0 & 0 & 0 & 0 & \gamma\lambda_1 & 0 &  0 & 0 & 0 &  0 & 0 & 0 & \lambda_1 & \lambda_2 & \lambda_3 & 0 & 0 & 0 & 0 &  0 & 0 & 0 & \gamma\lambda_5 & 0\\
 0 & 0 & 0 &  0 & 0 & 0 & 0 & \gamma\lambda_1^2 & 0 &  0 & 0 & 0 &  0 & 0 & 0 & \lambda_1^2 & \lambda_2^2 & \lambda_3^2 & 0 & 0 & 0 & 0 &  0 & 0 & 0 & \gamma\lambda_5^2 & 0\\
\tikzmark{left} 0 & 0 & 0 &  0 & 0 & 0 &0 &  0 & \gamma & 0 & 0 & 0 &  0 & 0 & 0 &0 &  0 & \gamma &  0 & 0 & 0 &  0 & 0 & 0 &0 &  0 & \gamma\\
 0 & 0 & 0 &  0 & 0 & 0 &0 &  0 & \gamma\lambda_1 & 0 & 0 & 0 &  0 & 0 & 0 &0 &  0 & \gamma\lambda_2 &  0 & 0 & 0 &  0 & 0 & 0 &0 &  0 & \gamma\lambda_5\\
 0 & 0 & 0 &  0 & 0 & 0 &0 &  0 & \gamma\lambda_1^2 & 0 & 0 & 0 &  0 & 0 & 0 &0 &  0 & \gamma\lambda_2^2 &  0 & 0 & 0 &  0 & 0 & 0 &0 &  0 & \gamma\lambda_5^2 \tikzmark{right} \end{array} \right]
\DrawBox[black]
\left[ \begin{array}{c} 
x_0\\x_1\\x_2\\x_3\\x_4\\x_5\\x_6\\x_7\\x_8\\x_9\\x_{10}\\x_{11}\\x_{12}\\x_{13}\\x_{14}\\x_{15}\\x_{16}\\
x_{17}\\x_{18}\\x_{19}\\x_{20}\\x_{21}\\x_{22}\\x_{23}\\x_{24}\\x_{25}\\x_{26}
\end{array} \right]}
=0.
\end{aligned}
\end{equation}
\vspace*{0.3in}
\begin{equation}\label{eq:tilde}
\begin{aligned}
\footnotesize{\left[ 
\begin{array}{*{21}{@{\hspace*{.05in}}c}}
 \tikzmark{left}1 & 1  & 0 & 0 & 0 & 0 & 0 &  1 & 0 & 0 & 0 & 0 & 0 & 0 & 1 & 0 & 0 & 0 & 0 & 0 & 0 \\
\lambda_1 & \lambda_2 &  0 & 0 & 0 & 0 & 0 & \lambda_2 & 0 & 0 & 0 & 0 & 0 & 0 & \lambda_5 & 0 & 0 & 0 & 0 & 0 & 0 \\
\lambda_1^2 & \lambda_2^2 & 0 & 0 & 0 & 0 & 0 & \lambda_2^2  & 0 & 0 & 0 & 0 & 0 & 0 & \lambda_5^2 & 0 & 0 & 0 & 0 & 0 & 0 \\
0 & \gamma  & 0 & 0 & 0 & 0 & 0 & 1 & 1 &  0 & 0 & 0 & 0 & 0 & 0 & 1 & 0 & 0 & 0 & 0 & 0 \\
0 & \gamma\lambda_1  & 0 & 0 & 0 & 0 & 0 & \lambda_1 & \lambda_2 & 0 & 0 & 0 & 0 & 0 & 0 & \lambda_5 & 0 & 0 & 0 & 0 & 0 \\
0 & \gamma\lambda_1^2 &  0 & 0 & 0 & 0 & 0 & \lambda_1^2 & \lambda_2^2 & 0 & 0 & 0 & 0 & 0 & 0 & \lambda_5^2 & 0 & 0 & 0 & 0 & 0 \tikzmark{right}\\
\DrawBox[black]
 0 & 0  &  1 & 1 & 1 & 0 & 0 &  0 & 0 &  1 & 0 & 0 & 0 & 0 & 1 &  0 &  1 & 0 & 0 & 1 & 0 \\
 0 & 0  & \lambda_1 & \lambda_2 & \lambda_3 & 0 & 0 &  0 & 0 & \lambda_2 & 0 & 0 & 0 & 0 & \lambda_4 & 0 & \lambda_5 & 0 & 0 & \lambda_6 & 0 \\
 0 & 0  & \lambda_1^2 & \lambda_2^2 & \lambda_3^2 & 0 & 0 &  0 & 0 & \lambda_2^2 & 0 & 0 & 0 & 0 & \lambda_4^2 & 0 & \lambda_5^2 & 0 & 0 & \lambda_6^2 & 0\\
 0 & 0  & 0 & \gamma & 0 & 0 & 0  &  0 & 0 & 1 & 1 & 1 & 0 & 0 & 0 &  1 & 0 & 1 & 0 & 0 & 1\\
 0 & 0  & 0 & \gamma\lambda_1 & 0 & 0 & 0 &  0 & 0 & \lambda_1 & \lambda_2 & \lambda_3 & 0 & 0 & 0 & \lambda_4 & 0 & \lambda_5 & 0 & 0 & \lambda_6 \\
 0 & 0  & 0 & \gamma\lambda_1^2  & 0 & 0 & 0 &  0 & 0 & \lambda_1^2 & \lambda_2^2  & \lambda_3^2 & 0 & 0 & 0 &  \lambda_4^2  & 0 &  \lambda_5^2  & 0  & 0 & \lambda_6^2\\
\tikzmark{left} 0 & 0  & 0 &  0 & \gamma & 0 & 0 &  0 & 0 & 0 &  0 & \gamma & 0 & 0 & 0 &  0 & 0 & 0 &  1 & 0 & 0 \\
 0 & 0  & 0 &  0 & \gamma\lambda_1 & 0 & 0 &  0 & 0 & 0 &  0 & \gamma\lambda_2 & 0 & 0 & 0 &  0 & 0 &  0 & \lambda_5 & 0 &  0 \\
 0 & 0  & 0 &  0 & \gamma\lambda_1^2 & 0 & 0 &  0 & 0 & 0 &  0 & \gamma\lambda_2^2 & 0 & 0 & 0 & 0 & 0 &  0 & \lambda_5^2 & 0 & 0 \\
 0 & 0 &  0 & 0 & 0 & 1 & 1 &  0 & 0 & 0 & 0 & 0 & 1 & 0 & 0 & 0 & 0 & 0 & 0 & \gamma & 0 \\
 0 & 0  &  0 & 0 & 0 & \lambda_1 & \lambda_2 &  0 & 0 & 0 & 0 & 0 & \lambda_2 & 0 & 0 & 0 & 0 & 0 & 0 & \gamma\lambda_5 & 0\\
 0 & 0 &  0 & 0 & 0 & \lambda_1^2 & \lambda_2^2 &  0 & 0 & 0 & 0 & 0 & \lambda_2^2 & 0 & 0 &  0 & 0 & 0 & 0 & \gamma\lambda_5^2 & 0\\
 0 & 0 &  0 & 0 & 0 & 0 & \gamma  &  0 & 0 & 0 & 0 & 0 & 1 & 1 & 0 & 0 & 0 & 0 &  0 & 0 & \gamma\\
 0 & 0 &  0 & 0 & 0 & 0 & \gamma\lambda_1 &  0 & 0 & 0 & 0 & 0 & \lambda_1 & \lambda_2 & 0 & 0 & 0 & 0 &  0 & 0 & \gamma\lambda_5 \\
 0 & 0 &  0 & 0 & 0 & 0 & \gamma\lambda_1^2 &  0 & 0 &  0 & 0 & 0 & \lambda_1^2 & \lambda_2^2 & 0 & 0 & 0 & 0 &  0 & 0 & \gamma\lambda_5^2 \tikzmark{right}  \end{array} \right] 
\DrawBox[black]
\left[ \begin{array}{c} 
x_0\\
x_1\\
x_3\\
x_4\\
x_5\\
x_6\\
x_7\\
x_9\\
x_{10}\\
x_{12}\\
x_{13}\\
x_{14}\\
x_{15}\\
x_{16}\\
x_{18}\\
x_{19}\\
x_{21}\\
x_{22}\\
x_{23}\\
x_{24}\\
x_{25}\end{array} \right]
 =0.
}
\end{aligned}
\end{equation}
\end{figure*}

Since the coefficients $\lambda_i$ are distinct for different $i$, 
the highlighted rows in \eqref{eq:huge}  imply that  $x_2=x_8=x_{11}=x_{17}=x_{20}=x_{26}=0.$ 
Eliminating these variables from \eqref{eq:huge}, we obtain a system of equations given by \eqref{eq:tilde}.

Looking at the first three rows in \eqref{eq:tilde}, and treating $x_1+x_9$ as a new variable, we conclude that
$x_0=x_1+x_9=x_{18}=0.$ Similarly, the second group of three rows implies that 
$\gamma x_1+x_9=x_{10}=x_{19}=0.$ Taking these results together and noting that $\gamma\ne 1,$ we see that
$x_0=x_1=x_{18}=x_9=x_{10}=x_{19}=0.$


%


A similar argument used for the last 9 rows in \eqref{eq:tilde} shows that
$x_5=x_{14}=x_{23}=x_6=x_7+x_{15}=x_{24}=
\gamma x_7+x_{15}=x_{16}=x_{25}=0,$ and so
$x_5=x_{14}=x_{23}=x_6=x_7=x_{24}=x_{15}=x_{16}=x_{25}=0.$ 
Writing out the remaining equations, we obtain the following set of equations:
$$
\left[ \begin{array}{*{6}{c}}
1 & 1 & 1 & 0 & 1 & 0\\
\lambda_1 & \lambda_2 & \lambda_2 & 0 & \lambda_5 & 0\\
\lambda_1^2 & \lambda_2^2 & \lambda_2^2 & 0 & \lambda_5^2 & 0\\
0 & \gamma & 1 & 1 & 0 & 1\\
0 & \gamma\lambda_1 & \lambda_1 & \lambda_2 & 0 & \lambda_5\\
0 & \gamma\lambda_1^2 & \lambda_1^2 & \lambda_2^2 & 0 & \lambda_5^2
\end{array}\right]
\left[\begin{array}{c}
x_3\\
x_4\\
x_{12}\\
x_{13}\\
x_{21}\\
x_{22}
\end{array}\right]=0.
$$
From this set of equations we obtain that $x_4+x_{12}=x_3=x_{21}=\gamma x_4+x_{12}=x_{13}=x_{22}=0.$ Thus,
$x_3=x_4=x_{21}=x_{12}=x_{13}=x_{22}=0.$ Overall these arguments prove that $X=0,$ and so $B$ is invertible. \qed

\subsection{A proof of the MDS property}
Let us fix $s,r,$ and $m$, so $n=sm$ and $l=s^m.$ The code $\sC$ given by Construction \ref{construction:cns} is an MDS array code
if for any $1\le i_1<i_2<\dots<i_r\le n,$ the matrix
\begin{multline*}
 B_{s,r,m}[i_1,i_2,\dots,i_r] \\ 
:= 
 \left[\begin{array}{cccc}
A_{0,i_1} & A_{0,i_2} & \dots & A_{0,i_r}\\
A_{1,i_1} & A_{1,i_2} & \dots & A_{1,i_r}\\
\vdots & \vdots & \vdots & \vdots\\
A_{r-1,i_1} & A_{r-1,i_2} & \dots & A_{r-1,i_r}\\
\end{array}\right]
\end{multline*}
is invertible. Below we suppress the parameters $s,r,m,$ and $i_1,i_2,\dots,i_r$ from the notation and write $B$ to refer to this matrix. In other words, given a vector $X\in F^{rl}$ we need to prove that
    \begin{equation}\label{eq:general}
      BX=0
    \end{equation}
 implies that $X=0,$ where $X$ is a vector in $F^{rl}$ with coordinates $X=(x_0, x_1, \dots, x_{rl-1})^T.$ 
 The proof essentially follows the example in Sect.~\ref{subsect:example}. We begin with a preview which also serves to
 introduce some notation.
 
 \textcolor{black}{In order to transform $B$ into a matrix of the form \eqref{eq:huge}, let us
define a permutation matrix $(P_{ij})_{0\le i,j < rl}$ by setting 
     \begin{equation}\label{eq:defP}
     P_{ij}=1\text{ iff } i=(j-j\,\text{mod\,}l)/l+r (j\,\text{mod\,}l);
     \end{equation}
     compare with our example in \eqref{eq:mod9}. }

Define a matrix $D=PB.$ We shall prove that
\begin{equation}\label{eq:dgen}
DX=0
\end{equation}
implies that $X=0.$
(The full notation for $D$ should be $D_{s,r,m}[i_1,i_2,\dots,i_r],$ but we again suppress the parameters.) For $a=0,1,\dots,l-1,$ define $D^{(a)}$ to be the $r\times rl$ submatrix of $D$ consisting of rows $ar,ar+1,\dots,ar+r-1.$ 
Define a column vector
\begin{equation}\label{eq:defL}
L_i=(1,\lambda_i,\dots,\lambda_i^{r-1})^T, \quad i\in[n].
\end{equation}

On account of \eqref{eq:osp}, for every $a\in\{0,1,\dots,l-1\},$ all the nonzero columns of $D^{(a)}$ belong to the set 
$$
\{L_1,L_2,\dots,L_n,\gamma L_1,\gamma L_2,\dots,\gamma L_n\}.
$$
We proceed by defining several subsets of the set of column indices of $D^{(a)}$ \textcolor{black}{for every $a\in\{0,1,\dots,l-1\}$}:
\begin{itemize}
\item Let $\sU^{(a)}\subset[n]$ be a subset such that $i\in \sU^{(a)}$ if and only if  there is a nonzero column in $D^{(a)}$ equal
to either $L_i$ or $\gamma L_i;$
\item Let $\sJ^{(a)}\subset\{0,1,\dots,rl\}$ be the set of indices of the nonzero columns in $D^{(a)}$;
\item \textcolor{black}{For $i\in[n]$,} define the set $\sJ^{(a)}(i)\subset\sJ^{(a)}$ as  $\sJ^{(a)}(i)=\{j\in\sJ^{(a)}: \text{ the $j$th column of $D^{(a)}$ is either $L_i$ or $\gamma L_i$} \}.$
\end{itemize}

In our example above, let $D^{(0)}$ be the first $r=3$ rows of the matrix $D=D_{3,3,2}[1,2,5]$ in \eqref{eq:huge}. Then the only nonzero
columns in $D^{(0)}$ are of type $L_1,L_2,L_3,$ or $L_5$, and so $\sU^{(0)}=\{1,2,3,5\}.$ The indices of the nonzero columns are given
by $\sJ^{(0)}=\{0,1,2,9,18\}$, and $\sJ^{(0)}(1)=\{0\}, \sJ^{(0)}(2)=\{1,9\}, \sJ^{(0)}(3)=\{2\},\sJ^{(0)}(5)=\{18\}.$

Clearly the sets $\sJ^{(a)}(i)$ form a partition of the set $\sJ^{(a)},$ so
    $$
      \sJ^{(a)}=\bigcup_{i\in\sU^{(a)}} \sJ^{(a)}(i).
    $$

\textcolor{black}{According to \eqref{eq:osp}, for every $i\in[n]$, every diagonal entry of $A_{t,i}$ is either 
$\lambda_i^t$ or $\gamma\lambda_i^t$ (see also the example in Sect.~\ref{subsect:example} where we explicitly write out the matrices $A_{t,1},\dots,A_{t,6}$).  
Therefore, for every $i\in[n]$, every row of $A_{t,i}$ contains at least one of the elements $\lambda_i^t$ and $\gamma\lambda_i^t$. 
As an immediate consequence, for every $i\in\{i_1,\dots,i_r\},$ the set of nonzero columns in the strip of $r$ rows $D^{(a)},a=0,1,\dots,l-1$ contains at least one column out of the pair $(L_i,\gamma L_i).$ }
This implies that
$\{i_1,i_2,\dots,i_r\}\subseteq\sU^{(a)}$ for all $a\in\{0,1,\dots,l-1\}.$
Our strategy of proving that \eqref{eq:dgen} is satisfied only for $X=0$ will be to find a set of indices 
    $$
    \sS=\{a:a\in\{0,1,\dots,l-1\},|\sU^{(a)}|= r\}
    $$
(as before $\sS=\sS_{s,r,m}[i_1,i_2,\dots,i_r]$).
\textcolor{black}{For every $a\in \sS$, all the nonzero columns of $D^{(a)}$ belong to the set 
$$
\{L_{i_1},L_{i_2},\dots,L_{i_r},\gamma L_{i_1},\gamma L_{i_2},\dots,\gamma L_{i_r}\}.
$$
Since the columns $L_{i_1},L_{i_2},\dots,L_{i_r}$ form a Vandermonde matrix, we conclude that the corresponding variables or 
their linear combinations are $0$, and therefore we can eliminate some of the variables in \eqref{eq:dgen}.}
Referring to our example, this set is exactly the set of highlighted rows in the matrix in \eqref{eq:huge}, and thus in this case
the set of strip labels equals $\cS=\{2,8\}.$ 

Suppose that such a set $\sS$ is found. Then the equations
    $$
D^{(a)}X=0,\quad  a\in\cS,
    $$
will imply that 
    \begin{equation}\label{eq:xt=0}
x_j=0 \text{ for all } j\in \bigcup_{a\in\cS} \sJ^{(a)}.
   \end{equation}
Using \eqref{eq:xt=0}, we can eliminate some of the variables in \eqref{eq:dgen} and obtain the system
   $$
   \widetilde{D}\widetilde{X}=0
   $$
(in the example this corresponds to obtaining \eqref{eq:tilde} from \eqref{eq:huge}).

Let $\tilde{l}=l-|\cS|$ be the remaining count of variables $x_i,$ so that $\widetilde{D}$ is an $r\tilde{l}\times r\tilde{l}$ matrix and $\widetilde{X}\in F^{r\tilde{l}}.$ We iterate the above steps and define subsets
  $
  \widetilde{D}^{(a)},\widetilde{\sU}^{(a)}
  $
 for all $a\in\{0,1,\dots,\tilde{l}-1\}.$ In the example the matrix $\widetilde{D}$ is given in \eqref{eq:tilde}, and
the new set $\widetilde \sS$ of the groups of $r$ equations is given by $\widetilde\sS= \{0,1,4,5,6\}.$ Restricting our
attention to the equations in these groups, we eliminate another subset of variables by proving that they are necessarily equal to zero, and continue this procedure until finally all of the variables have been shown to be zero. 
 

A rigorous proof uses induction and is given below.

\begin{theorem}\label{thm:MDS}
The code $\sC$ given by Construction \ref{construction:cns} is an  $(n=sm,k=n-r,l=s^m)$ MDS array code.
\end{theorem}
To prove this theorem we need to show that for every choice of the indices $1\le i_1<i_2<\dots<i_r\le n,$ the only $X$ that satisfies \eqref{eq:dgen} is the all-zero vector. This will follow from the next two lemmas.

\begin{lemma}\label{claim2}
For any $1\le i_1<i_2<\dots<i_r\le n,$
$$
\min_{a\in\{0,1,\dots,l-1\}} |\sU^{(a)}[i_1,i_2,\dots,i_r]|=r.
$$
\end{lemma}
\begin{IEEEproof}
As mentioned above, $\{i_1,i_2,\dots,i_r\}\subseteq\sU^{(a)}$  for any $a\in\{0,1,\dots,l-1\},$ so $\min_a|\sU^{(a)}|\ge r$ (we again simplify the notation by dropping the indices $i_1,\dots,i_r$).
We claim that there always exists an index $a\in\{0,1,\dots,l-1\}$ such that $\sU^{(a)}=\{i_1,i_2,\dots,i_r\}.$ 
To see this, define the (possibly empty) set
  \begin{equation}\label{eq:G}
\begin{aligned}
  \sG & =\{ v: v\in[m], \\
           & \{(v-1)s+1,(v-1)s+2,\dots,vs\}\subseteq\{i_1,i_2,\dots,i_r\}\}.
\end{aligned}
  \end{equation}
Now choose $a\in\{0,1,\dots,l-1\}$ such that 
     \begin{equation}\label{eq:[m]}
\text{if }v\in[m]\backslash\sG \text{ then } (v-1)s+a_v+1\notin \{i_1,i_2,\dots,i_r\}.
     \end{equation}
\textcolor{black}{To see that we can always find such an $a$, notice that by \eqref{eq:G}, for every 
$v\in[m]\backslash\sG$, there exists a number $y_v\in\{1,2,\dots,s\}$ such that 
$(v-1)s+y_v\notin \{i_1,i_2,\dots,i_r\}$. In order to satisfy \eqref{eq:[m]}, it suffices to set the $v$-th digit of $a$ to be $y_v-1$, i.e., to set $a_v=y_v-1$.}

\textcolor{black}{Next we prove that $\sU^{(a)}=\{i_1,i_2,\dots,i_r\}$, which is equivalent to the following statement:}

\textcolor{black}{
{\bf Claim:}
For any $i\in\{i_1,i_2,\dots,i_r\}$, all the nonzero entries of the $a$-th row of $A_{t,i}$ belong to the set 
$$
\{\lambda_{i_1}^t, \lambda_{i_2}^t, \dots, \lambda_{i_r}^t,
\gamma\lambda_{i_1}^t, \gamma\lambda_{i_2}^t, \dots, \gamma\lambda_{i_r}^t\}.
$$ }

\textcolor{black}{
Let us write $i=(v-1)s+u+1$, where $v\in[m]$ and $u\in\{0,1,\dots,s-1\}$.
We consider the following two cases:}

\vspace*{.05in}{\bf Case 1: $v\in[m]\backslash\sG$}

\textcolor{black}{ In this case \eqref{eq:[m]} states that $(v-1)s+a_v+1\notin \{i_1,i_2,\dots,i_r\},$ and therefore $(v-1)s+a_v+1\neq i$ (keep in mind that $i\in\{i_1,i_2,\dots,i_r\}$). As a result, $a_v \neq u$. 
According to \eqref{eq:osp}, if $a_v<u$, then the $a$-th row of $A_{t,i}$ contains a single nonzero entry $\lambda_i^t$; if $a_v>u$, then the $a$-th row of $A_{t,i}$ contains a single nonzero entry $\gamma\lambda_i^t$. Both $\lambda_i^t$ and $\gamma\lambda_i^t$ belong to the set $\{\lambda_{i_1}^t, \lambda_{i_2}^t, \dots, \lambda_{i_r}^t,
\gamma\lambda_{i_1}^t, \gamma\lambda_{i_2}^t, \dots, \gamma\lambda_{i_r}^t\}$. This establishes the claim for this case.}

\vspace*{.05in}{\bf Case 2: $v\in\sG$}

\textcolor{black}{If $a_v\neq u$, then the claim holds by the argument in Case 1, so let
$a_v=u.$ The $a$-th row of $A_{t,i}$ contains $s$ nonzero entries
$\lambda_{(v-1)s+1}^t,\lambda_{(v-1)s+2}^t,\dots,\lambda_{vs}^t$. By \eqref{eq:G}, they all belong to $\{\lambda_{i_1}^t, \lambda_{i_2}^t, \dots, \lambda_{i_r}^t\}$.
This establishes the claim for this case and completes the proof of the lemma.}

\remove{Suppose that there exists an element in $\sU^{(a)}$ different from $i_1,\dots,i_r.$ More specifically, suppose that
$v\in[m]$ and $u\in\{0,1,\dots,s-1\}$ are such that $(v-1)s+u+1\in\sU^{(a)}\backslash\{i_1,i_2,\dots,i_r\}.$
On account of \eqref{eq:osp}, $v$ has the property that $(v-1)s+a_v+1\in \{i_1,i_2,\dots,i_r\}.$
Therefore, by \eqref{eq:[m]}, $v\in\sG$, but then by \eqref{eq:G} we obtain that
$(v-1)s+u+1\in\{i_1,i_2,\dots,i_r\},$ a contradiction. 
Thus our assumption is false, and $\sU^{(a)}=\{i_1,i_2,\dots,i_r\}.$}
\end{IEEEproof}

\begin{lemma}\label{claim1}
\textcolor{black}{For every $a\in\{0,1,\dots,l-1\}$,}
$x_j=0$ for all $j\in \sJ^{(a)}.$
\end{lemma}
\begin{IEEEproof} 
We will argue by induction on the cardinality of the set $\sU^{(a)}.$ By Lemma \ref{claim2}, to establish the induction basis
we need to prove that the lemma holds for all $a$ such that $|\sU^{(a)}|=r.$ 
Let $a$ be one of the values that have this property. 
We will prove that
for every $t\in[r]$, $x_j=0$ for all $j\in\sJ^{(a)}(i_t).$

Let us write the index $i_t,t\in[r]$ in the form 
$$
i_t=(v_t-1)s+u_t+1,
$$
 where $v_t\in[m]$ and $0\le u_t\le s-1.$ 
Let us further partition $[r]$ into three disjoint subsets $\sK_1,\sK_2,\sK_3$ as follows:
    \begin{align*}
\sK_1 = & \{t:t\in[r],a_{v_t}=u_t\} \\
 & \cup\{t:t\in[r],/\!\!\!\exists\, p\in[r] \text{ s.t. } v_p=v_t \text{ and } u_p=a_{v_t}\},\\
\sK_2= & \{t:t\in[r],a_{v_t}>u_t\}\\
 & \cap\{t:t\in[r],\exists\, p\in[r] \text{ s.t. } v_p=v_t \text{ and } u_p=a_{v_t}\},\\
\sK_3 = & \{t:t\in[r],a_{v_t}<u_t\}\\
 & \cap\{t:t\in[r],\exists\, p\in[r] \text{ s.t. } v_p=v_t \text{ and } u_p=a_{v_t}\}.
   \end{align*}
We will prove our claim separately for each of these subsets, starting with $\sK_1.$ 
\textcolor{black}{By definition \eqref{eq:osp}, all the nonzero entries in the matrix $A_{{h},(v-1)s+u+1}$
belong to the set $\{\lambda_{(v-1)s+1}^{h}, \lambda_{(v-1)s+2}^{h}, \dots, \lambda_{vs}^{h},
\gamma\lambda_{(v-1)s+1}^{h}, \gamma\lambda_{(v-1)s+2}^{h}, \linebreak[4]\dots, \gamma\lambda_{vs}^{h}\}$, where ${h}=0,1,\dots,r-1.$
Moreover, if $a_v \neq u$, then the $a$-th row of the matrix $A_{{h},(v-1)s+u+1}$ contains only a single nonzero entry, either 
$\lambda_{(v-1)s+u+1}^{h}$ or $\gamma\lambda_{(v-1)s+u+1}^{h}$. On the other hand, if $a_v=u$, then the $a$-th row of the matrix $A_{{h},(v-1)s+u+1}$ contains $s$ nonzero entries. In particular, if $a_v=u$, then the only appearance of the element $\lambda_{(v-1)s+u+1}^{h}$ in the $a$-th row of all the matrices $A_{h,i}, i\in[n]$ is in position $(a,a)$ of  $A_{{h},(v-1)s+u+1}$, i.e., on the diagonal
of  $A_{{h},(v-1)s+u+1}$ (and $\gamma \lambda_{(v-1)s+u+1}^{h}$ does not appear in the $a$-th row of any one of the matrices $A_{h,i}, i\in[n]$). As an immediate consequence, if $a_v=u$, then the column $L_{(v-1)s+u+1}$ appears at most once
in the matrix $D^{(a)}$ (it appears when $(v-1)s+u+1\in\{i_1,\dots,i_r\}$), and $\gamma L_{(v-1)s+u+1}$ does not appear in $D^{(a)}.$
Therefore,} for each $t\in \sK_1$, we have $\sJ^{(a)}(i_t)=\{(t-1)l+a\}.$ The vectors $L_{i_j}, j=1,\dots,r$ are
linearly independent, which implies that $x_{(t-1)l+a}=0, t\in \sK_1.$

Moving to the case $t\in \sK_2,$ observe that $\sJ^{(a)}(i_t)=\{(t-1)l+a,(p-1)l+a(v_t,u_t)\}$ which implies that
    \begin{equation}\label{eq:gamma}
\gamma x_{(t-1)l+a}+x_{(p-1)l+a(v_t,u_t)}=0.
    \end{equation}  
Let us consider the submatrix $D^{(a(v_t,u_t))}.$ Its nonzero columns are described as follows:
   \begin{align*}
   \sU^{(a(v_t,u_t))} & =\{i_1,i_2,\dots,i_r\}, \\
   \sJ^{(a(v_t,u_t))}(i_p) & =\{(t-1)l+a,(p-1)l+a(v_t,u_t)\}.
   \end{align*}
From this we see that 
    \begin{equation}\label{eq:nogamma}
     x_{(t-1)l+a}+x_{(p-1)l+a(v_t,u_t)}=0.
   \end{equation}
Since $\gamma\ne1,$ conditions \eqref{eq:gamma} and \eqref{eq:nogamma} imply that $x_{(t-1)l+a}=x_{(p-1)l+a(v_t,u_t)}=0,$
exhausting the case of $t\in \sK_2.$

The last remaining case $t\in\sK_3$ is very similar to $\sK_2,$ and we again obtain that
$x_j=0$ for all $j\in \sJ^{(a)}(i_t).$ This concludes the proof of the induction basis.

Now suppose that the statement of the lemma holds true for all $a$ such that 
$|\sU^{(a)}|\le w-1$ for some $w\ge r$ and let us prove it for all $a$ such that $|\sU^{(a)}|=w.$
We again aim to show that for every $i\in \sU^{(a)},$
$x_j=0$ for all $j\in\sJ^{(a)}(i).$

For $i\in\sU^{(a)}\backslash\{i_1,i_2,\dots,i_r\},$ there exists a unique $t\in[r]$ such that $a_{v_t}=u_t$ and 
$1\le i-(v_t-1)s\le s,$ and $\sJ^{(a)}(i)=\{(t-1)l+a(v_t,\alpha)\},$ where $\alpha=i-(v_t-1)s-1.$ Consider the
matrix $D^{(a(v_t,\alpha))}.$ By our choice of $i$ there is no $p\in[r]$ such that $v_p=v_t$ and $u_p=\alpha.$
Therefore,
   $
   \sU^{(a(v_t,\alpha))}\subset\sU^{(a)}$
and $i\not\in \sU^{(a(v_t,\alpha))}.$
This implies that $|\sU^{(a(v_t,\alpha))}|\le w-1,$ so the induction hypothesis applies, and $x_j=0$ for all
$j\in\sJ^{(a(v_t,\alpha))}.$ Furthermore, $(t-1)l+a(v_t,\alpha)\in \sJ^{(a(v_t,\alpha))},$ and thus
$x_{(t-1)l+a(v_t,\alpha)}=0.$ Rephrasing this, we have shown that for every $i\in\sU^{(a)}\backslash\{i_1,i_2,\dots,i_r\}$,
$x_j=0$ for all $j\in\sJ^{(a)}(i).$ 

We are left to consider the variables $\cup_{t\in[r]}\{x_j:j\in \sJ^{(a)}(i_t)\}.$ To show that they must be 0 to satisfy $DX=0,$ we note that the left-hand side of $D^{(a)}X=0$ reduces to a linear combination of the linearly independent columns $L_{i_j},j=1,\dots,r$.
Therefore, the coefficients of this linear combination are all 0. This implies that for every $t\in[r]$, $x_j=0$ for all
$j\in \sJ^{(a)}(i_t)$. This claim is proved in exactly the same way as the induction basis above, so we omit the proof. 
This completes the induction step.
\end{IEEEproof}

\begin{corollary}
The code $\sC'$ given by Construction \ref{construction:ext} is an $(n=rm+r',k=n-r,l=r^{m+1})$ MDS array code.
\end{corollary}
\begin{IEEEproof} Consider the MDS code $\sC$ of length $n=r(m+1)$ given by Construction \ref{construction:cns}.
The code $\sC'$ is obtained from $\sC$ by discarding $r-r'$ coordinates, and so is also MDS.
\end{IEEEproof}

{We finish this section by a brief remark on the complexity of node repair (decoding) and encoding of the
constructed codes. From the coding-theoretic
perspective, the repair problem is erasure correction, and is very similar to the encoding problem (the encoding is correction
of $r=n-k$ erasures from $k$ known nodes).}
From the example in Section~\ref{subsect:example} and the proof of Theorem \ref{thm:MDS}, we immediately see that the encoding and decoding procedures of codes given by either of Constructions \ref{construction:cns} or \ref{construction:ext}
rely on inversion of $r\times r$ matrices over $F$, and thus have low complexity.

\section{the optimal access property}\label{sect:opac}

Consider the code $\sC$ given by Construction \ref{construction:cns}.
\textcolor{black}{\textcolor{black}{Recall that we write the $i$-th node as 
$C_i=(c_{i,0}, c_{i,1},\dots, c_{i,l-1})^T.$} 
For every $i\in[n]$,}
define the following set of coordinates of the $i$-th node: 
    \begin{equation}\label{eq:C}
    \cC_i^{(v,u)}=\{c_{i,a}:a\in\{0,1,\dots,l-1\},a_v=u\}.
    \end{equation}
where $a_v$ is the $v$-th digit of $a.$
\begin{theorem}\label{thm:opac}
Consider the code $\sC$ given by Construction \ref{construction:cns}.
For any $v\in[m]$ and $u\in\{0,1,\dots,s-1\},$ the node $C_{(v-1)s+u+1}$ can be recovered from the elements in the set
$$
\cC^{(v,u)}=\bigcup_{\begin{substack}{i\in[n]\\i\neq (v-1)s+u+1}\end{substack}}\cC_i^{(v,u)}.
$$
\end{theorem}
\begin{IEEEproof} Fix a value of $t$.
Let us write out the $a$-th row of \textcolor{black}{the} equation $\sum_{i=1}^n A_{t,i}C_i=0$ 
(cf.~\eqref{eq:parityform}). We first notice that since $n=sm$, the equation 
$\sum_{i=1}^n A_{t,i}C_i=0$  is equivalent to
$\sum_{q=1}^m \sum_{w=0}^{s-1} A_{t,(q-1)s+w+1}C_{(q-1)s+w+1}=0$.
By definition \eqref{eq:osp}, if $a_q>w$, then the $a$-th row of $A_{t,(q-1)s+w+1}$ contains a single nonzero entry $\gamma \lambda_{(q-1)s+w+1}^t$ located in the $a$-th column; if $a_q<w$, then the $a$-th row of $A_{t,(q-1)s+w+1}$ contains a single nonzero entry $\lambda_{(q-1)s+w+1}^t$ located in the $a$-th column; if $a_q=w$, then the $a$-th row of $A_{t,(q-1)s+w+1}$ contains $s$ nonzero entries located in columns $a(q,0)$ to $a(q,s-1)$.
Thus, the $a$-th row of the equation $\sum_{i=1}^n A_{t,i}C_i=0$ can be written as follows:
\begin{multline}\label{eq:aq}
\sum_{q\in[m]}
\Big( \sum_{w=0}^{a_q-1}\gamma \lambda_{(q-1)s+w+1}^t c_{(q-1)s+w+1,a} \\  +\sum_{w=0}^{s-1} \lambda_{(q-1)s+w+1}^t c_{(q-1)s+a_q+1,a(q,w)} \\
 +\sum_{w=a_q+1}^{s-1} \lambda_{(q-1)s+w+1}^t c_{(q-1)s+w+1,a}\Big) = 0,
\end{multline}
where the first sum in the parentheses corresponds to the case $a_q>w$; the second sum corresponds to the case $a_q=w$; and the third sum  corresponds to the case $a_q<w$.
Since our aim is to repair the node $C_{(v-1)s+u+1}$, let us break the sum on $q\in[m]$ in \eqref{eq:aq} into two parts: $q\neq v$ and $q=v$. We obtain the  equation given in \eqref{eq:optac}.
\begin{figure*}
\begin{equation}\label{eq:optac}
\begin{aligned}
\sum_{q\neq v,q\in[m]}
\Big( \sum_{w=0}^{a_q-1}\gamma \lambda_{(q-1)s+w+1}^t c_{(q-1)s+w+1,a}
+\sum_{w=0}^{s-1} \lambda_{(q-1)s+w+1}^t c_{(q-1)s+a_q+1,a(q,w)}
 +\sum_{w=a_q+1}^{s-1} \lambda_{(q-1)s+w+1}^t c_{(q-1)s+w+1,a}\Big)\\
+\Big( \sum_{w=0}^{a_v-1}\gamma \lambda_{(v-1)s+w+1}^t c_{(v-1)s+w+1,a}+\underline{\sum_{w=0}^{s-1} \lambda_{(v-1)s+w+1}^t c_{(v-1)s+a_v+1,a(v,w)}}
 +\sum_{w=a_v+1}^{s-1} \lambda_{(v-1)s+w+1}^t c_{(v-1)s+w+1,a}\Big)=0
\end{aligned}
\end{equation}
\end{figure*}
For all $t=0,1,\dots,r-1$ and all $a$ satisfying $a_v=u,$ all the terms in \eqref{eq:optac} apart from the underlined term can be found from the elements in the set
$\cC^{(v,u)}.$
Indeed, \textcolor{black}{on account of \eqref{eq:np} and the fact that}
\begin{figure*}
\begin{equation}\label{eq:np}
\left[\begin{array}{c}
\sum_{w=0}^{s-1} c_{(v-1)s+u+1,a(v,w)}\\
\sum_{w=0}^{s-1} \lambda_{(v-1)s+w+1} c_{(v-1)s+u+1,a(v,w)}\\
\vdots\\
\sum_{w=0}^{s-1} \lambda_{(v-1)s+w+1}^{r-1} c_{(v-1)s+u+1,a(v,w)}
\end{array}\right]=
\left[\begin{array}{cccc}
1 & 1 & \dots & 1\\
\lambda_{(v-1)s+1} & \lambda_{(v-1)s+2} & \dots & \lambda_{(v-1)s+s}\\
\vdots & \vdots & \vdots & \vdots\\
\lambda_{(v-1)s+1}^{r-1} & \lambda_{(v-1)s+2}^{r-1} & \dots & \lambda_{(v-1)s+s}^{r-1}
\end{array}\right]
\left[\begin{array}{c}
c_{(v-1)s+u+1,a(v,0)}\\
c_{(v-1)s+u+1,a(v,1)}\\
\vdots\\
c_{(v-1)s+u+1,a(v,s-1)}
\end{array}\right].
\end{equation}
\end{figure*}
 $r\ge s,$ the coordinates in the set $\{c_{(v-1)s+u+1,a(v,w)}:w=0,1,\dots,s-1\}$ can be found from the set
$\{\sum_{w=0}^{s-1} \lambda_{(v-1)s+w+1}^t c_{(v-1)s+u+1,a(v,w)}:t=0,1,\dots,r-1\}$
for all $a=0,1,\dots,l-1.$ In particular, the values in the set
  \begin{align*}
& \{c_{(v-1)s+u+1,a}:a=0,1,\dots,l-1\} \\
= & \{c_{(v-1)s+u+1,a(v,w)}:a_v=u,w=0,1,\dots,s-1\}
  \end{align*}
can be found from the values
  $
  \Big\{\sum_{w=0}^{s-1} \lambda_{(v-1)s+w+1}^t c_{(v-1)s+u+1,a(v,w)}:a_v=u,t=0,1,\dots,r-1\Big\}.
  $
As mentioned above, the values $\{\sum_{w=0}^{s-1} \lambda_{(v-1)s+w+1}^t c_{(v-1)s+u+1,a(v,w)}:a_v=u,t=0,1,\dots,r-1\}$ 
are uniquely determined by  the elements in the set $\cC^{(v,u)}.$
We conclude that the entire node $C_{(v-1)s+u+1}$ can be determined by the elements in $\cC^{(v,u)}.$
\end{IEEEproof}

\begin{theorem} Let $s=r,$ then the code $\sC$ given by Construction \ref{construction:cns} has the optimal access property.
\end{theorem}
\begin{IEEEproof}
Since $\cC_i^{(v,u)}$ contains exactly a {$(1/r)$th} fraction of the coordinates of $C_i,$ 
by Theorem \ref{thm:opac} we only need to access a {$1/r$ proportion} of the data stored in each of the 
surviving nodes in order to repair a single node failure. 
\end{IEEEproof}

\begin{theorem}
The code $\sC'$ given by Construction \ref{construction:ext} has the optimal access property.
\end{theorem}
\begin{IEEEproof}
Using the same approach as in Theorem \ref{thm:opac}, we can show that 
for all $v\in[m]$ and $u\in\{0,1,\dots,r-1\},$ the node $C_{(v-1)r+u+1}$ can be determined by the elements in the set $\{c_{i,a}:i\neq (v-1)r+u
+1,a_v=u\},$ and that for all $u\in\{0,1,\dots,r'-1\},$ the node $C_{mr+u+1}$ can be determined by the elements in the set $\{c_{i,a}:i\neq mr+u
+1,a_{m+1}=u\}.$ 
\end{IEEEproof}

Note that upon setting $s=r$ in Construction \ref{construction:cns}, we obtain optimal-access MDS array codes with code length divisible by $r,$ and Construction \ref{construction:ext} gives optimal-access MDS array codes with code length not divisible by $r.$ Thus we can construct optimal-access $(n,n-r,r^{\lceil n/r\rceil})$ MDS array codes for any $n$ and $r.$

\subsection{Group optimal access} In the second part of this section we examine the group optimal access property of
the codes considered above and prove the following result.
\begin{theorem}\label{thm:goa}
The code $\sC$ given by Construction \ref{construction:cns} has the $(s,s+k-1)$-group optimal access property.
\end{theorem}
This theorem will follow from Theorem~\ref{thm:opac} and a lemma proved below in this section.

Let us define the following subset of indices
  $$
  \sN^{(v)}=[n]\backslash\{(v-1)s+1,(v-1)s+2,\dots,vs\}, \quad v\in[m].
  $$
  
\begin{lemma}\label{thm:setMDS}
Consider the code $\sC$ given by Construction \ref{construction:cns}.
For every $v\in[m]$ and $u\in\{0,1,\dots,s-1\},$
and every $\sM\subseteq \sN^{(v)}$ with cardinality $|\sM|=k,$
the values of elements in the set $\cup_{i\in\sN^{(v)}} \cC_i^{(v,u)}$ $($cf. \eqref{eq:C}$)$
are determined by the elements in the set $\cup_{i\in\sM} \cC_i^{(v,u)}.$
\end{lemma}
\textcolor{black}{Before proving Lemma~\ref{thm:setMDS}, let us explain how this lemma together with
Theorem~\ref{thm:opac} implies Theorem~\ref{thm:goa}.
The group optimal access property simply means that for every $v\in[m]$ and every $u\in\{0,1,\dots,s-1\}$, the node $C_{(v-1)s+u+1}$ can be repaired by connecting to 
$\{C_{(v-1)s+u'+1}:u'\in\{0,1,\dots,s-1\}\setminus\{u\}\}$
together with any other $k$ helper nodes in the set $\{C_i:i\in \sN^{(v)}\},$
and accessing exactly $(1/s)$th fraction of coordinates of each helper node.
In Theorem~\ref{thm:opac}, we have shown that the node $C_{(v-1)s+u+1}$ can be repaired if we know the values of all the elements in the set
$\bigcup_{i\neq (v-1)s+u+1}\cC_i^{(v,u)}$
from all the surviving nodes. By definition each set $\cC_i^{(v,u)}$ contains exactly {$(1/s)$th} fraction of the coordinates of $C_i$. This is where we need Lemma~\ref{thm:setMDS} which states that the values of the elements in the set $\cup_{i\in\sN^{(v)}} \cC_i^{(v,u)}$ $($cf. \eqref{eq:C}$)$
can be calculated from the elements in the set $\cup_{i\in\sM} \cC_i^{(v,u)}$ for any $\sM\subseteq \sN^{(v)}$ such that $|\sM|=k.$ This establishes that the code $\sC$ given by Construction \ref{construction:cns} has the $(s,s+k-1)$-group optimal access property. }

\begin{IEEEproof} (of Lemma~\ref{thm:setMDS})
The case $s=r$ is trivially true, so we assume that $s<r.$ Let us write \eqref{eq:parityform} in matrix form:
\begin{equation}\label{eq:defM}
\left[\begin{array}{cccc}
A_{0,1} & A_{0,2} & \dots & A_{0,n}\\
A_{1,1} & A_{1,2} & \dots & A_{1,n}\\
\vdots & \vdots & \vdots & \vdots \\
A_{r-1,1} & A_{r-1,2} & \dots & A_{r-1,n}
\end{array}\right]
\left[\begin{array}{c}
C_1\\
C_2\\
\vdots\\
C_n
\end{array}\right]=0.
\end{equation}
Let us permute the equations in this system using the permutation matrix defined in \eqref{eq:defP} and denote the resulting
matrix of coefficients by $H$. As before, let $H^{(a)}, a=0,1,\dots,l-1$ be a submatrix of $H$ formed of rows
$ar,ar+1,\dots,(a+1)r-1.$
\textcolor{black}{Writing out the equation
$H^{(a)}[C_1,C_2,\dots,C_n]^T=0$
in full form, we obtain \eqref{eq:Ha},}
\begin{figure*}
\begin{equation}\label{eq:Ha}
\begin{aligned}
  \sum_{q\in[m]\backslash v} \Big(\sum_{w=0}^{a_q-1}\gamma c_{(q-1)s+w+1,a} L_{(q-1)s+w+1} 
+\sum_{w=0}^{s-1} c_{(q-1)s+a_q+1,a(q,w)} L_{(q-1)s+w+1} 
      +\sum_{w=a_q+1}^{s-1} c_{(q-1)s+w+1,a} L_{(q-1)s+w+1} \Big)\\
=-\Big(\sum_{w=0}^{a_v-1}\gamma c_{(v-1)s+w+1,a} L_{(v-1)s+w+1} 
+\sum_{w=0}^{s-1} c_{(v-1)s+a_v+1,a(v,w)} L_{(v-1)s+w+1}
  +\sum_{w=a_v+1}^{s-1} c_{(v-1)s+w+1,a} L_{(v-1)s+w+1} \Big)
\end{aligned}
\end{equation}
\end{figure*}
where the vectors $L_i,i\in[n]$ are defined in \eqref{eq:defL} (this equation amounts to taking columns 
$(v-1)s+1,(v-1)s+2,\dots,vs$ to the right-hand side of \eqref{eq:defM}).

Define polynomials $g_0^{(v)}(x)=\prod_{w=1}^s (x-\lambda_{(v-1)s+w}),$
and $g_j^{(v)}(x)=x^j g_0^{(v)}(x)$ for $j=0,1,\dots,r-s-1.$
Since the degree of $g_j^{(v)}(x)$ is less than $r$ for all $j=0,1,\dots,r-s-1,$ we can write
    $$
g_j^{(v)}(x)=\sum_{t=0}^{r-1}g_{j,t}^{(v)}x^t.
    $$
Define the $(r-s)\times r$ matrix
$$
G^{(v)}=\left[\begin{array}{cccc}
g_{0,0}^{(v)} & g_{0,1}^{(v)} & \dots & g_{0,r-1}^{(v)} \\
g_{1,0}^{(v)} & g_{1,1}^{(v)} & \dots & g_{1,r-1}^{(v)} \\
\vdots & \vdots & \vdots & \vdots\\
g_{r-s-1,0}^{(v)} & g_{r-s-1,1}^{(v)} & \dots & g_{r-s-1,r-1}^{(v)}
\end{array}\right].
$$
We have
\begin{multline}\label{eq:pL}
 G^{(v)}L_i \\
=  
\left[\begin{array}{cccc}
g_{0,0}^{(v)} & g_{0,1}^{(v)} & \dots & g_{0,r-1}^{(v)} \\
g_{1,0}^{(v)} & g_{1,1}^{(v)} & \dots & g_{1,r-1}^{(v)} \\
\vdots & \vdots & \vdots & \vdots\\
g_{r-s-1,0}^{(v)} & g_{r-s-1,1}^{(v)} & \dots & g_{r-s-1,r-1}^{(v)}
\end{array}\right]
\left[\begin{array}{c}
1\\
\lambda_i\\
\vdots\\
\lambda_i^{r-1}
\end{array}\right] \\
=\hat{L}_i^{(v)},
\end{multline}
where $\hat{L}_i^{(v)},i\in[n]$ is given by
\begin{equation}\label{eq:hatL}
\hat{L}_i^{(v)}=
\left[\begin{array}{c}
g_0^{(v)}(\lambda_i)\\
g_1^{(v)}(\lambda_i)\\
\vdots\\
g_{r-s-1}^{(v)}(\lambda_i)
\end{array}\right]=g_0^{(v)}(\lambda_i) 
\left[\begin{array}{c}
1\\
\lambda_i\\
\vdots\\
\lambda_i^{r-s-1}
\end{array}\right].
\end{equation}
By definition, $g_0^{(v)}(\lambda_i)=0$ for all $(v-1)s+1\le i\le vs,$ and so
$G^{(v)}L_i=0$ for all $(v-1)s+1\le i\le vs.$
Observe that every term on the right-hand-side of \eqref{eq:Ha} contains one column vector from 
the set $\{L_i:(v-1)s+1\le i\le vs\}.$ As a result, multiplying equation \eqref{eq:Ha} by $G^{(v)}$ on the left, 
we obtain
\begin{multline*}
\sum_{q\neq v,q\in[m]}
\Big(\sum_{w=0}^{a_q-1}\gamma c_{(q-1)s+w+1,a} G^{(v)}L_{(q-1)s+w+1} \\
+\sum_{w=0}^{s-1} c_{(q-1)s+a_q+1,a(q,w)} G^{(v)}L_{(q-1)s+w+1} \\
+\sum_{w=a_q+1}^{s-1} c_{(q-1)s+w+1,a} G^{(v)}L_{(q-1)s+w+1} \Big)=0.  
\end{multline*}
Using \eqref{eq:pL}, this equation can be written as
\begin{multline}\label{eq:hat}
\sum_{q\neq v,q\in[m]}
\Big(\sum_{w=0}^{a_q-1}\gamma  c_{(q-1)s+w+1,a} \hat{L}_{(q-1)s+w+1}^{(v)} \\
+\sum_{w=0}^{s-1} c_{(q-1)s+a_q+1,a(q,w)}  \hat{L}_{(q-1)s+w+1}^{(v)} \\
+\sum_{w=a_q+1}^{s-1} c_{(q-1)s+w+1,a}  \hat{L}_{(q-1)s+w+1}^{(v)} \Big)=0.
\end{multline}
In order to prove the theorem, we only need to prove that
given any $v\in[m]$ and $u\in\{0,1,\dots,s-1\},$
and any $i_1<i_2<\dots<i_{r-s}$ such that $\{i_1,i_2,\dots,i_{r-s}\}\subseteq \sN^{(v)},$
the values of elements in the set $\cup_{t=1}^{r-s} \cC_{i_t}^{(v,u)}$ 
can be determined by the elements in the set $\cup_{i\in\sM} \cC_i^{(v,u)},$ where
$\sM=\sN^{(v)}\backslash\{i_1,i_2,\dots,i_{r-s}\}$. 
We will prove that we can find the elements in the set $\cup_{t=1}^{r-s} \cC_{i_t}^{(v,u)}$ from  $\cup_{i\in\sM} \cC_i^{(v,u)}$ using equation \eqref{eq:hat}.

For a given $a=0,1,\dots,l-1,$ denote by $E^{(a)}$ the set of equations in \eqref{eq:hat}. Each set $E^{(a)}$ 
contains $r-s$ scalar equations. Observe that 
if $a_v=u,$ then $E^{(a)}$ contains only elements in $\cup_{i\in\cN^{(v)}} \cC_i^{(v,u)}.$
Moreover, every element in the set $\cup_{i\in\cN^{(v)}} \cC_i^{(v,u)}$ appears at least once in the equations $\{E^{(a)}:a_v=u\}.$
Therefore,  equations \eqref{eq:hat} contain $(r-s)l/s$ scalar equations and 
$(r-s)l/s$ unknown elements, namely, the elements in the set $\cup_{t=1}^{r-s} \cC_{i_t}^{(v,u)}.$

Let us set all the elements in the set $\cup_{i\in\cM} \cC_i^{(v,u)}$ to 0 in $E^{(a)}$ and denote by $E_{\sM}^{(a)}$
 the  obtained set of equations.
(In other words, $E_{\cM}^{(a)}$ are the equations obtained by eliminating all the terms which contain elements in the set
$\cup_{i\in\cM} \cC_i^{(v,u)}$ in \eqref{eq:hat}.)
In order to prove the lemma, it suffices to show that the equations
$\{E_{\sM}^{(a)}:a_v=u\}$
imply that all the elements in the set
$\cup_{t=1}^{r-s} \cC_{i_t}^{(v,u)}$ are $0.$

Recall that equation \eqref{eq:dgen} can be viewed as the set of equations 
$\{D^{(a)}[i_1,i_2,\dots,i_r]X=0:a\in\{0,1,\dots,l-1\}\}.$
Note that, once we form a vector $X$ of the elements in the set $\cup_{t=1}^{r-s} \cC_{i_t}^{(v,u)},$ equations
$\{E_{\sM}^{(a)}:a_v=u\}$ have almost the same form as $\{D^{(a)}[i_1,i_2,\dots,i_r]X=0:a\in\{0,1,\dots,l-1\}\}.$ 
The only difference is that in the equations $\{E_{\sM}^{(a)}:a_v=u\},$ the columns 
$\{\hat{L}_i^{(v)}:i\in\sN^{(v)}\}$ take place of the vectors $\{L_i:i\in[n]\},$ and $r-s$ takes place of $r.$

Examining closely the proof of Theorem \ref{thm:MDS}, we note that the property that any $r$ columns in 
the set $\{L_i,i\in[n]\}$ are linearly independent, suffices to show that $X=0.$ 
Since $p_0^{(v)}(\lambda_i)\neq 0$ for any $i\in\sN^{(v)},$ by Definition \eqref{eq:hatL}
any $(r-s)$ vectors in the set of vectors $\{\hat{L}_i^{(v)}:i\in\sN^{(v)}\}$ are also linearly independent.
Thus, using exactly the same arguments as in the proof of Theorem \ref{thm:MDS}, we can show that
the equations
$\{E_{\sM}^{(a)}:a_v=u\}$ imply that all the elements in the set
$\cup_{t=1}^{r-s} \cC_{i_t}^{(v,u)}$ are $0.$
This completes the proof of the theorem.
\end{IEEEproof}

\section{Conclusion}
\textcolor{black}{
In this paper we presented an explicit construction of optimal-access MDS codes with nearly optimal sub-packetization $l=r^{\lceil n/r \rceil}$ and field size $|F|\ge r\lceil n/r \rceil,$ which is just slightly greater than $n$.
It is shown in \cite{Wang14} that if we only require systematic optimal repair property instead of optimal access property, there
exist codes with even
smaller sub-packetization $l=r^{\lceil k/(r+1) \rceil}$ if the base field $F$ is  sufficiently large. For $r=2$ and $3$, explicit constructions achieving this sub-packetization over fields of size linear in $k$ are given in \cite{Wang14} and \cite{Raviv15}, respectively. It is an interesting open problem to give explicit constructions achieving this sub-packetization over small fields for general values of $r$.}

\end{document}